\documentclass[prb,aps,twocolumn,superscriptaddress,floatfix]{revtex4-1}
\usepackage[english]{babel}
\usepackage[dvipsnames]{xcolor}
\usepackage{amsmath} 
\usepackage{color}
\usepackage{graphicx}
\usepackage{dcolumn}
\usepackage{bm}
\usepackage[colorlinks=true, linkcolor=blue, citecolor=blue, urlcolor=blue]{hyperref}
\usepackage{amssymb} 

\begin{document}

 \title{
Deterministic roughening in the dc-driven precessional regime of domain walls
 }

\author{E. F. Pusiol}
\affiliation{Centro At\'omico Bariloche, CNEA-CONICET, 
(R8402AGP) San Carlos de Bariloche, Río Negro, Argentina}

\author{V. Lecomte}
\affiliation{
Laboratoire Interdisciplinaire de Physique, Université Grenoble Alpes \& CNRS (UMR 5588), 140 avenue de la Physique, 38402 Saint-Martin d'Heres, France
}

\author{S. Bustingorry}
\affiliation{Centro At\'omico Bariloche, Instituto de Nanociencia y Nanotecnología, CNEA-CONICET, 
(R8402AGP) San Carlos de Bariloche, Río Negro, Argentina}

\author{A. B. Kolton}
\affiliation{Centro At\'omico Bariloche, CNEA-CONICET, 
(R8402AGP) San Carlos de Bariloche, Río Negro, Argentina}
\affiliation{Instituto Balseiro, UNCUYO,  
(R8402AGP) San Carlos de Bariloche, Río Negro, Argentina}

\date{\today}

\begin{abstract}
We numerically study the dynamics of extended domain walls in homogeneous ferromagnets driven by a uniform magnetic field at zero temperature. Using both micromagnetic Landau–Lifshitz–Gilbert simulations and a collective-coordinate description, we show that flat chiral domain walls become linearly unstable above the Walker breakdown field and below a higher threshold, provided their length exceeds a characteristic scale. This instability is captured by a quasi-universal spectral stability diagram, parameterized solely by the Gilbert damping, which predicts the onset of deviations from rigid-wall behavior. Beyond the linear regime, large domain walls with bands of unstable modes develop spatiotemporal chaos, intricate Bloch-line dynamics, and deterministic roughening. At a critical field, the system undergoes a dynamical phase transition from a flat to a rough moving phase with universal features. Our results provide a framework for addressing domain-wall dynamics in the presence of thermal fluctuations and quenched disorder by disentangling their effects from intrinsic deterministic instabilities.\end{abstract}

\pacs{Valid PACS appear here}
\maketitle

\section{Introduction}
The dynamics of magnetic domain walls (DWs) driven by external magnetic fields or electric currents is key to understanding technologically relevant properties of ferromagnetic materials \cite{Parkin2008,allwood2005magnetic,kumar2022domain,Venkat_2024}. Beyond applications, extended DWs also raise fundamental questions in nonequilibrium physics. Theoretical progress has come from modeling DWs as elastic objects with no internal structure~\cite{chauve2000creep,ferrero2021creep}, where even weak disorder leads to rich universal behavior~\cite{lemerle1998domain,Jeudy2016,Grassi2018,pardo2017,Albornoz2021,kolton2023,Durin2024}. As we show here, however, intrinsic nonlinear effects, present even without disorder, can profoundly shape DW dynamics, requiring explicit treatment of internal degrees of freedom.

Incorporating internal degrees of freedom into DW dynamics poses a significant theoretical challenge, even without disorder or thermal fluctuations. A key feature is the Walker field, a critical value of the applied field predicted for perfectly flat 180° DWs~\cite{Schryer1974,hubert1998magnetic}, which separates a stationary regime (with fixed internal magnetization and increasing velocity) from a precessional regime (with rotating magnetization and oscillatory motion). In the latter, the mean DW velocity shows negative differential mobility just above the Walker field~\cite{beach2005dynamics,Dourlat2008}. While this anomaly is not problematic for particle-like DWs 
it leads to an intrinsic instability in extended DWs at zero temperature in homogeneous media. This gives rise to a corrugated phase~\cite{Slonczewski1972} marked by complex dynamics and spontaneous Bloch line nucleation.
Internal degrees of freedom thus become essential at finite velocities, where rigid-wall approximations fail.

While the Walker breakdown has been widely studied for particle-like DWs~\cite{Schryer1974,thiaville2005,hillebrands2006spin,Mougin2007,dipietro2023}, extended DWs in thin ferromagnetic films under constant magnetic fields have received less attention. A key advance was linking velocity anomalies to linearly unstable flexural modes~\cite{thevenard2011domain,gourdon2013domain, Kim2024}, but prior studies focused on small systems with few unstable modes and weak nonlinear coupling. Consequently, the impact of nonlinearity on the large-scale roughness, velocity, and long-time dynamics of extended DWs remains poorly understood.

In this work, we study the dynamics of extended DWs with perpendicular magnetic anisotropy using a reduced DW model~\cite{malozemoff1979magnetic}, validated against micromagnetic Landau–Lifshitz–Gilbert (LLG) simulations. We compute a quasi-universal instability diagram, dependent only on Gilbert damping, that predicts mode instabilities as a function of applied field. This reveals a \textit{dynamical phase transition}
from a corrugated, spatiotemporally chaotic achiral state to a flat chiral moving phase at a critical driving field. Our framework provides a basis for understanding these nonlinear effects across varying system sizes and field strengths. Furthermore, our results highlight this system as a rich example of nonequilibrium pattern formation and roughening emerging entirely in the absence of disorder or thermal fluctuations.

\section{Models and Methods}

Our base model for the DW dynamics driven by a constant magnetic field is the LLG equation
\begin{equation}
\dfrac{\partial\mathbf{m}}{\partial t}= -\gamma\mathbf{m}\times\mathbf{H}_{\text{eff}}+\alpha\mathbf{m}\times\dfrac{\partial\mathbf{m}}{\partial t}
\label{eq_LLG}
\end{equation}
where $\gamma$ is the reduced gyromagnetic ratio and $\alpha$ the Gilbert damping. Using spherical coordinates for the unit magnetization vector
$\mathbf{m} = (\sin\theta \cos\varphi, \sin\theta \sin\varphi, \cos\theta)$, 
the effective field  $\mathbf{H}_{\text{eff}}=-(1/\mu_0M_s) \delta E/\delta \mathbf{m}$ is derived from the total magnetic energy functional
\begin{align}
    \label{eq:energy}
    E &= \int \Bigl\{  A_\text{ex}[(\nabla\theta)^2+\sin^{2}\theta(\nabla\varphi)^2]-K_u\cos^2\theta  \\
    &-\mu_0H_zM_s\cos\theta +\dfrac{1}{2}\mu_0M_s^{2} [N_n(\mathbf{m} \cdot \mathbf{n})^{2}+\cos^{2}\theta] \Bigl\}d\mathbf{r}
    \nonumber
\end{align}
where $M_s$ is the saturation magnetization, $A_\text{ex}$ is the exchange stiffness constant, $K_u$ the anisotropy constant, $\mu_0$ the magnetic permeability, and $N_n$ is the demagnetizing factor along the DW normal direction~\cite{skaugen2019}. We will assume a constant field applied along the $z-$axis, of strength $H_z$ and a directed DW with its average normal in the direction of the unit vector ${\bf n}\equiv \hat y$. 
The last term in Eq.~\eqref{eq:energy} is a local thin-film approximation to magnetostatic effects, where the $\cos^2\theta$ contribution accounts for the dominant out-of-plane shape anisotropy, while the $N_n(\mathbf m\!\cdot\!\mathbf n)^2$ term represents the weaker in-plane demagnetizing energy associated with a directed domain wall.

To gain further insight, we adopt a reduced collective--coordinate description of the DW dynamics, commonly referred to as the $u$--$\phi$ model, which captures the most relevant degrees of freedom while remaining analytically transparent and computationally efficient. In the present work, we formulate this model in a convenient nondimensional form that enables a systematic exploration of stability diagrams and long--time spatiotemporal dynamics, including chaotic regimes that would be prohibitively costly to access using the full LLG equation alone. We assess the validity of the reduced description by direct comparison with zero--temperature LLG simulations in a homogeneous, isotropic thin--film ferromagnet, focusing on one--dimensional domain wall dynamics.

\begin{figure}[t]
    \centering
    \includegraphics[width=\linewidth]{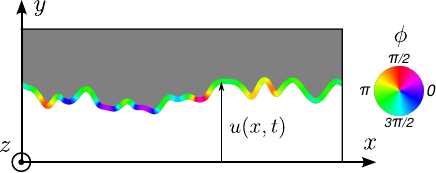}
    \caption{
Snapshot of a 
DW 
configuration 
highlighting its collective coordinates $u(x,t)$ and $\phi(x,t)$.
    }
\label{fig:snap}
\end{figure}

\subsection{$u$--$\phi$ Model}

The reduced collective-coordinates model is derived by substituting a Bloch domain-wall ansatz
\begin{align}
    \theta(x, y, t) &= 2 \arctan\left( \exp[{(y - u(x, t))}/{\Delta}] \right),     \label{eq:Bloch1} \\
    \label{eq:Bloch2}
    \varphi(x, y, t) &= \phi(x, t), 
\end{align}
into the LLG equations, where $u(x, t)$ and $\phi(x, t)$ represent the local position and the internal magnetization angle of the wall, respectively. This derivation assumes a constant DW width $\Delta = \sqrt{A_\text{ex}/\left(K_u - \mu_0 M_s^2/2\right)}$, where the denominator accounts for the effective anisotropy including the magnetostatic demagnetizing energy. Furthermore, the DW is treated as a single-valued, directed interface (see Fig. \ref{fig:snap}), precluding the formation of overhangs or pinch-off loops. 

To simplify the analysis, we nondimensionalize the model using characteristic length, time, and magnetic-ﬁeld scales (See Appendix \ref{App:uphi_deriv} for details). Distances are measured in units of $u_0 = \Delta$ along the $y$-direction, and $L_0 = \Delta ({2\tilde{K}/N_n})^{1/2}$ in the $x$-direction, where $\tilde{K} = 2 K_u / (\mu_0 M_s^2) - 1$. Time is measured in units of $T_0 =({1+\alpha^{2}})/({\gamma M_{s} N_{n}})$, and the applied field $h$ is expressed in units of the Walker breakdown field\cite{skaugen2019} $H_W = \alpha M_s N_n / 2$, so $h=1$ in the model corresponds exactly to the nondimensionalized Walker field by construction. The resulting equations constitute the $u-\phi$ or Slonczewski model~\cite{Slonczewski1972}:
 \begin{align}
\dot{u} &= \tfrac{1}{2} \left[ \alpha^2 h + \sin(2\phi) + \alpha \partial_x^2 u - \partial_x^2 \phi \right], \nonumber \\
\dot{\phi} &= \tfrac{1}{2} \left[ \alpha h - \alpha \sin(2\phi) + \partial_x^2 u + \alpha \partial_x^2 \phi \right].
\label{eq:uphi}
\end{align}
To study extended DWs and their collective dynamics, we numerically solve the full set of coupled nonlinear equations using a pseudo-spectral Crank–Nicolson scheme (See Appendix \ref{App:uphi_num}). Starting from a weakly perturbed flat DW of uniform chirality, we track its geometric and transport properties over time across various system sizes.

\subsection{Micromagnetic Simulations}
\label{sect:mmag}

Since the $u$--$\phi$ model described by Eqs.~\eqref{eq:uphi} relies on several approximations, we validate its predictions against full micromagnetic simulations. 
To this end we perform LLG simulations using the Mumax³ software package~\cite{vansteenkiste2014design}. The simulated thin film has dimensions $L$ in the $x$ direction (with periodic boundary conditions) and a sufficiently large dimension in the $y$ direction to ensure that edge effects are negligible and effectively mimic an infinite space for domain wall dynamics. The simulation volume was discretized using a uniform grid with cell dimensions of $2 \times 2 \times 0.5$ nm³. The moving window method provided by the solver was utilized to dynamically maintain the DW at the centre of the film throughout the simulation.

As an example of a quasi-two-dimensional system, we simulate a Co/Pt multilayer with the following parameters~\cite{Metaxas2007,skaugen2019,skaugen2022}: $\alpha = 0.27$, $A_{\mathrm{ex}} = 1.4\times 10^{-11}$ J/m, $M_s = 9.1\times 10^{5}$ A/m, and $K_u = 8.4\times 10^{5}$ J/m³. The DW position $u(x,t)$ was determined by locating the zero-crossing of $m_{z}(y)$, obtained through linear interpolation between adjacent grid points with opposite magnetization.

At $t=0$, the magnetization angles $(\theta, \varphi)$ were set according to Eqs.~\eqref{eq:Bloch2}. We set $\phi(x,0)=\phi_{0}$, and the initial wall position $u(x,0)$ as a small-amplitude perturbation of the flat DW configuration. A non-zero perturbation is required since a perfectly flat DW remains planar during evolution, and no instability can develop. 

To bridge the $u$--$\phi$ model and LLG simulations, a renormalization of the system size is required. This is because the reduced model treats the DW width $\Delta$ as a material constant independent of film geometry, whereas in micromagnetic simulations, the effective width $\Delta'$ is modified by the finite film thickness. For instance, for the specific Co/Pt multilayer parameters used here, the reduced model predicts $\Delta \approx 6.62$~nm, while micromagnetic calculations yield a slightly narrower $\Delta' \approx 6.356$~nm. To achieve a consistent comparison for a system of size $L$ in the $u$--$\phi$ model, we scale the LLG simulation length to $L'=L(\Delta' / \Delta)$. This ensures that the dimensionless system length, $\tilde{L} = L/\Delta = L'/\Delta'$, remains identical in both frameworks (see also Appendix~\ref{app:matchingscales} for further discussions). A comprehensive analytical treatment of the thickness dependence of $\Delta$ and its impact on the $u$--$\phi$ mapping is detailed in Ref.~\cite{skaugen2019}.

\section{Linear stability of flat domain walls}
\label{sec:linearstabilitydiagrams}

Before presenting our linear stability analysis, we briefly review the existing understanding of DW instabilities. Heuristic arguments have long suggested that \textit{negative differential mobility} is associated with morphological instabilities of planar DWs~\cite{Slonczewski1972, krizakova2019, Garcia2021}. Consider a DW driven by an external field $h$, with average velocity \( V(h) \). Assuming weak local DW curvature and linear elasticity, the velocity becomes spatially modulated by the local curvature:
\begin{align}
    \frac{\partial u}{\partial t} 
    &\approx V(h+C \partial_x^2 u) \approx V(h) + C\, \partial_h V(h)\, \partial_x^2 u,
\end{align}
where \( C \propto  \sigma  \), with \( \sigma \) the DW tension and $h+C \partial_x^2 u$ representing a local effective field. Fourier-transforming in space, the evolution of each mode \( u_\kappa \) becomes
\begin{align}
    u_{\kappa\neq 0}(t) &\approx u_{\kappa\neq 0}(0)\, \exp\left[-C \kappa^2 \partial_h V(h)t\right].
\label{eq:heuristicinstability}
\end{align}
According to this heuristic argument the fate of the curvature modes then depends on the sign of the time-averaged differential mobility: if \( \partial_h V(h) > 0 \), perturbations (by imposing $u_{\kappa \neq 0}(0) \neq 0$) decay and the DW remains flat, otherwise \( \partial_h V(h) < 0 \), perturbations grow, indicating instability.
This heuristic argument links negative differential mobility of a rigid wall to corrugation instabilities. 
Specifically, it predicts instability in the range \( 1 < h < h_S \), with
\begin{equation}
h_{\mathrm{S}} = \frac{\alpha^2 + 1}{\alpha \sqrt{\alpha^2 + 2}},    
\label{eq:hslonczewski}
\end{equation}
corresponding to the local minimum of the rigid-wall velocity in the precessional regime.

Here, we develop a linear stability analysis based on the $u$--$\phi$ model [Eqs.~\eqref{eq:uphi}] that, in contrast to simplified treatments, explicitly accounts for the internal degrees of freedom of the DW structure in a mathematically rigorous manner. We linearize around the flat-wall solution,
\begin{align}
u(x, t) &= Q \left(x,t \right)+u_{1}(t), \\
\phi(x, t) &= \Phi\left(x,t \right)+\phi_{1}(t), 
\end{align}
where $u_1$ and $\phi_1$ obey the rigid-wall Walker equations:
\begin{align}
    \dot{u}_{1} &= \dfrac{1}{2}[{\alpha^{2} h + \sin(2\phi_{1})}], \\
\dot{\phi}_{1} &= \dfrac{1}{2}[{\alpha h - \alpha \sin( 2\phi_{1})}].
\end{align}
If $\Phi \ll \phi_1$ (we do \textit{not} require $Q \ll u_1$), $\sin(2\phi) \approx \sin(2\phi_1)+ 2 \cos(2\phi_1) \Phi$. Then, 
\begin{align}
    \partial_t Q &=  \cos(2\phi_1) \Phi + 
    \frac{1}{2}(\alpha Q''-\Phi'') 
    \\
\partial_t \Phi &= - \alpha \cos(2\phi_1) \Phi +
\frac{1}{2}(Q''+ \alpha \Phi''). 
\label{eq:lineardynamics}
\end{align}
To decouple the linear system we use spatial Fourier transform: $Q(x,t)\rightarrow{}\hat{Q}_\kappa(t)$, $\Phi(x,t)\rightarrow{}\hat{\Phi}_\kappa(t)$, where $\kappa$ denotes a mode wavevector in units of $1/L_{0}$. For each mode, we obtain a $2\times2$ linear system as follows:
\begin{align}
    \partial_{t} \hat{Q}_\kappa &=
\frac{\kappa^2}{2}  (\hat{\Phi}_\kappa -\alpha \hat{Q}_\kappa) +\hat{\Phi}_\kappa \cos(2\phi_{1}) \label{eq:fourier1}\\
    \partial_{t}  \hat{\Phi}_\kappa
&=
-\frac{\kappa^2}{2} (\hat{Q}_\kappa +\alpha \hat{\Phi}_\kappa) -\alpha\hat{\Phi}_\kappa \cos(2\phi_{1}).
\label{eq:fourier2}
\end{align}
For $h < 1$, in the so-called stationary regime, the system of Eq.~\eqref{eq:fourier2} is autonomous with negative eigenvalue real parts (see Appendix~\ref{sec:belowwalkerfield}), confirming stability of flat DWs. For $h \geq 1$, in the precessional regime, the system of Eq.~\eqref{eq:fourier2} is non-autonomous due to time-dependent $\phi_1$. Since $\dot{\phi}_1 > 0$ for $h > 1$, we reparametrize time via $\phi_1$ and rewrite the system of linearized equations in Fourier space as  
$
(\partial_{\phi_{1}}\hat{Q}_\kappa, \partial_{\phi_{1}}\hat{\Phi}_\kappa)^T = A_{\kappa}(\phi_{1})(\hat{Q}_\kappa, \hat{\Phi}_\kappa)^T,
$
with
\begin{equation}
A_{\kappa}(\phi_{1})=
    \begin{pmatrix}
        \dfrac{-\kappa^{2}}{h-\sin(2\phi_{1})} & \dfrac{-\kappa^{2}+2\cos(2\phi_{1})}{\alpha(h-\sin(2\phi_{1}))}\\[10pt]
        \dfrac{-\kappa^{2}}{\alpha(h-\sin(2\phi_{1}))} & \dfrac{-\kappa^{2}-2\cos(2\phi_{1})}{h-\sin(2\phi_{1})}
    \end{pmatrix},
    \label{eq:linearabove}
\end{equation}
the dynamical matrix.
Since $A_{\kappa}(\phi_1) = A_{\kappa}(\phi_1 + n\pi)$ for integer $n$, we can apply Floquet theory \cite{chicone2006ordinary}. The general solution can be written as:
\begin{equation}
    \begin{pmatrix}
        \hat{Q}_\kappa(\phi_1)\\
        \hat{\Phi}_\kappa(\phi_1)
    \end{pmatrix}
 = P(\phi_1) e^{R \phi_1}     
 \begin{pmatrix}
        \hat{Q}_\kappa(0)\\
        \hat{\Phi}_\kappa(0)
    \end{pmatrix},
    \label{eq:floquetsys}
\end{equation}
where \(P(\phi_1)\) is a periodic matrix with period \(\pi\)
and \(R\) is a constant matrix. 
The matrix $M_\kappa \equiv P(\pi) e^{R \pi}$ is the \textit{monodromy} matrix and can be computed by integrating the linear system Eq.~\eqref{eq:floquetsys} over $\phi_1 \in [0, \pi]$ with initial conditions $(\hat Q_{\kappa}(0), \hat \Phi_{\kappa}(0)) = (1, 0)$ and $(0, 1)$. 
The first solution yields the first column and the second solution the second column of $M_\kappa$.
Since
\begin{equation}
 \begin{pmatrix}
        \hat{Q}_\kappa(n \pi)\\
        \hat{\Phi}_\kappa(n \pi)
    \end{pmatrix}    
= M_\kappa^n 
\begin{pmatrix}
        \hat{Q}_\kappa(0)\\
        \hat{\Phi}_\kappa(0)
    \end{pmatrix},    
\end{equation}
a flat DW is unstable for mode $\kappa$ if the largest eigenvalue $\mu_{\kappa}$ of $M_\kappa$, or Floquet multiplier, satisfy $|\mu_{\kappa}| > 1$, while the real and imaginary parts of $\mu_{\kappa}$ yield separately additional details on how modes either grow or decay as a function of the number of periods of $\phi_1$.

Floquet multipliers can be efficiently computed numerically, in parallel, since all $\kappa$ are decoupled. Additionally, the reparametrization of time $t \to \phi_1$ is convenient as it allow us to freeze the integration interval to compute $M_\kappa$ for different $h$, changing the integral over the physical $h$-dependent period proportional to $\sqrt{h^2-1}$ for a dimensionless integral over $\pi$. 

It is worth noting that thanks to our non-dimensionalization, it is evident that Floquet multipliers depend on $\alpha$, $\kappa$ and $h$. Therefore, the stability map determined by them is \textit{quasi-universal} as it describes a family of systems with the same $\alpha$, with their wave-vectors and applied fields related by simple field and length scale transformations, but otherwise independent of other microscopic details.

\section{Results}

\begin{figure}[t]
    \centering
    \includegraphics[width=\linewidth]{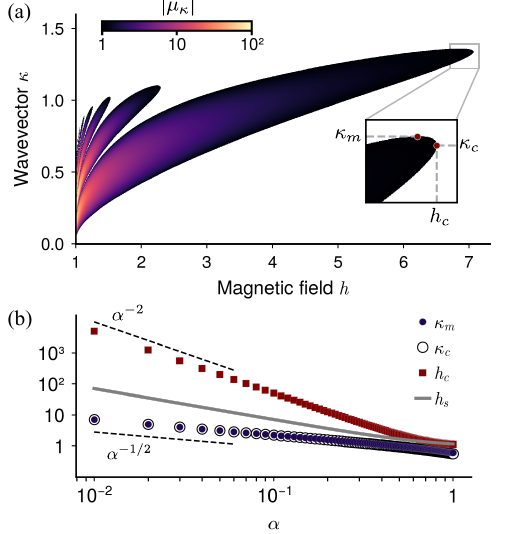}
    \caption{
    (a) Linear stability diagram of DW modes for $\alpha = 0.27$. White regions are stable; colors indicate the magnitude of the largest Floquet multiplier for unstable modes with $|\mu_\kappa| > 1$.
    (b) Dependence of characteristic wavenumbers $\kappa_m$, $\kappa_c$, and critical field $h_c$ (as defined in inset of panel a) on Gilbert damping $\alpha$.
    }
    \label{fig:stabilitydiag}
\end{figure}

Figure~\ref{fig:stabilitydiag}(a) shows a heat map of $|\mu_{\kappa}|$ as a function of $\kappa$ and $h$ for $\alpha = 0.27$, with unstable regions ($|\mu_\kappa|>1$) highlighted in color, with a color scale corresponding to the magnitude of the multiplier (black corresponding to the threshold case $|\mu_{\kappa}|=1$).
These linear instability diagrams are quasi-universal as they depend solely on the Gilbert damping $\alpha$, displaying a rich changing structure as $\alpha$ is varied over an experimentally relevant range (see Appendix~\ref{app:LS}).
In this sense, the choice $\alpha=0.27$ is not special, except that it enables direct comparison with previous numerical and experimental results for Co/Pt thin film multilayer systems~\cite{Metaxas2007,skaugen2019,skaugen2022}.

A flat DW becomes linearly unstable when any allowed mode falls within the “instability feathers”, which extend beyond the negative-mobility regime ($1<h<h_S$) of the rigid DW model. These quasi-universal stability maps depend only on $\alpha$, and can be applied to any system with the same damping by converting $(\kappa, h)$ to physical units using $L_0$ and the Walker field $H_W$.

The spectral instability is bounded by the maximum unstable wavevector $\kappa_m$ and a critical field $h_c$ above which flat DWs are stable. The point $(\kappa_c, h_c)$ in the inset of Fig.~\ref{fig:stabilitydiag}(a) marks a transition to a corrugated phase. For $\alpha \ll 1$, the scalings $\kappa_c \approx \kappa_m \sim \alpha^{-1/2}$ and $h_c \sim \alpha^{-2}$ [Fig.~\ref{fig:stabilitydiag}(b)] highlight the key role of Gilbert damping in setting the instability’s characteristic length and field scales and how stability diagrams stretch/compress in both directions with changing $\alpha$ (see Fig.~\ref{fig:gallery} in the Appendix).

\begin{figure}[t]
    \centering
    \includegraphics[width=\linewidth]{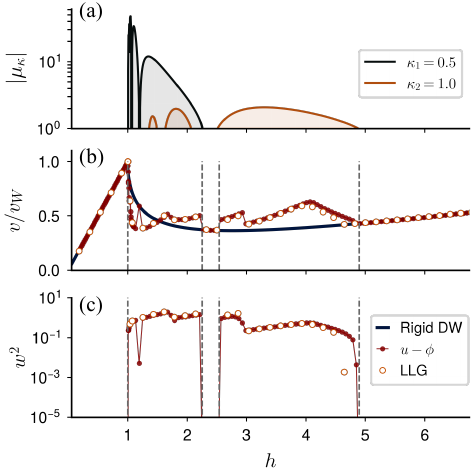}
    \caption{ 
        (a) Absolute value of Floquet multipliers $|\mu_{\kappa}|$ as a function of driving field $h$, for $\alpha=0.27$, showing the only two conditionally unstable modes of a small system, for $\kappa_{1}=0.5$ and $\kappa_{2}=2\kappa_{1}=1.0$. (b) Normalized mean DW velocity and (c) roughness obtained from $u$--$\phi$ model and micromagnetic simulations for the same parameters and length scales. For comparison, panel (b) also shows the rigid wall velocity $\langle \dot u_{1}\rangle$ (blue line).
    }
    \label{fig:finitesize}
\end{figure}

The linear stability map of Fig.~\ref{fig:stabilitydiag} sets the stage for exploring \textit{nonlinear} effects in the steady-state DW dynamics. In the $u$--$\phi$ model, nonlinearity arises solely from the $\sin(2\phi)$ term in Eqs.~\eqref{eq:uphi}, potentially triggering cascade-like interactions and spectral broadening. We first focus on small, yet experimentally relevant systems, where mode coupling is easier to analyze.

\subsection{Finite-size effect}
\label{sec:finitesize}

From Fig.~\ref{fig:stabilitydiag}(a) we predict that for $h > 1$, if the system size satisfies $L/L_0 < 2\pi/\kappa_m$, all nonzero modes remain linearly stable, and perturbations decay—implying rigid DW motion. When $L/L_0 > 2\pi/\kappa_m$ however, the fundamental mode can become unstable over some $h$ range, even if higher modes remain stable. These finite-size effects are general and predicted for a wide range of values of $\alpha$ [see Fig.~\ref{fig:gallery} in the Appendix, and Fig. \ref{fig:stabilitydiag}(b)].

As a concrete example, consider a DW of length $L$ such that $\kappa_{1}=2\pi/(L/L_{0})=0.5$. For the damping parameter $\alpha=0.27$ shown in Fig.~\ref{fig:stabilitydiag}(a), only two modes can became unstable for $h>1$: one associated with the fundamental mode $\kappa_{1}=0.5$ and another with the second mode ($\kappa_{2}=2\kappa_{1}=1.0)$. This is illustrated in Fig.~\ref{fig:finitesize}(a), which shows the absolute value of the Floquet multipliers $|\mu_{\kappa}|$ for both the fundamental and second modes as a function of $h$. From Fig.~\ref{fig:finitesize}(a) we thus predict instabilities in the $1<h \lesssim 5$ range, except for small windows inside this range where the two modes become simultaneously stable.

To investigate how the two linearly unstable modes of Fig.~\ref{fig:finitesize}(a) influence the full nonlinear steady-state dynamics and also to benchmark the $u$--$\phi$ model against LLG simulations, we performed steady-state runs using matched micromagnetic parameters (see Section~\ref{sect:mmag}). Both models were configured with an identical fundamental wavenumber $\kappa_{1}$. This ensures that the discrete set of available modes $\kappa_n = n \kappa_1$ is consistent across both frameworks, which requires a precise mapping of the $u$--$\phi$ dimensionless length scales to the physical dimensions of the micromagnetic simulation volume.

Figures~\ref{fig:finitesize}(b)--(c) compare the roughness \(w^2\) via the DW displacements
\begin{align}
w^2 = 
\langle u^2\rangle -
\langle u \rangle
^2 ,
\label{eq:roughness}
\end{align}
and the mean velocity,
\begin{equation}
    v = \left\langle \dfrac{\partial u}{\partial t} \right\rangle,
\end{equation}
normalized by the Walker velocity $v_W$, obtained from simulations with both models where $\langle \dots \rangle$ denotes steady-state average. For a stable flat DW, we expect $w^2 = 0$ and

\begin{equation}
  v = \langle \dot u_1 \rangle =
  \begin{cases}
    \dfrac{h}{2}({\alpha^2+1}) & \text{if} \;h\leq1\\\\
    \dfrac{h}{2}\alpha^2 - \dfrac{1}{2} \left(\sqrt{h^2-1}-h\right)      & \text{if } \;h>1
  \end{cases}
\end{equation}

As can be appreciated in 
Fig.~\ref{fig:finitesize}(b)-(c),
deviations from this rigid motion in both observables $w^2$ and $v$ fairly coincide with the unstable windows (i.e. the range of fields for which at least one mode is unstable) shown in Fig.~\ref{fig:finitesize}(a), linking linear instability to nonlinear steady states. In these unstable regimes, the DW develops a finite number of vertical Bloch lines (VBLs)~\cite{Slonczewski1972,hubert1998magnetic,Hutner2019} detected as localized \(\pi\)-kinks in \(\phi(x,t)\) (fast color changes in  Fig.~\ref{fig:snap}), while flat walls show uniform internal angle. A comprehensive analysis of the roughness and velocity as a function of system length $L$, showing how a different number of linear unstable modes affects $w^2$ and $v$ is presented in Appendix~\ref{sec:w2andvvshfinitesize}.

Notably, the simultaneous discontinuities in velocity and roughness at specific values of $h$ provide a stringent test of the reduced model's accuracy; the discrepancy in the critical fields for these transitions is less than $5\%$ (see Appendix \ref{app:matchingscales}, Fig. \ref{fig:matching}, for a similar discontinuity matching test but with a different system size $L$). The most pronounced deviations occur in regimes of weak linear instability, where the Floquet multipliers are only marginally greater than unity [see darker regions in Fig.~\ref{fig:stabilitydiag}(a)]. 
Overall, the results presented in Fig.~\ref{fig:finitesize}(b)-(c) validate the $u$--$\phi$ framework defined by Eqs.~\eqref{eq:uphi} across both stationary and precessional driving regimes.

Fig.~\ref{fig:finitesize} also demonstrates that domain-wall corrugation can extend significantly beyond the negative differential mobility regime, $1 < h < h_S$, extending to $h_c > h_S$ as anticipated by the linear stability analysis of Section~\ref{sec:linearstabilitydiagrams}. Specifically, a comparison between the velocity minimum $h_S$ of Eq.~\eqref{eq:hslonczewski} and the critical field $h_c$ in Fig.~\ref{fig:stabilitydiag}(b) reveals that the heuristic argument, reviewed in 
Sec.~\ref{sec:linearstabilitydiagrams} and leading to Eq.~\eqref{eq:heuristicinstability}, 
fails to predict the maximum field for corrugation, particularly for small values of $\alpha$, where $h_S \sim \alpha^{-1}$ and $h_c \sim \alpha^{-2}$. This finding also stands in contrast to the interpretation of moving corrugated phases made in previous studies~\cite{Slonczewski1972,khovenkil2003corrugation,krizakova2019,Garcia2021} which typically associate the instability solely with the region of negative differential mobility after the Walker breakdown.

In summary, results of Fig. \ref{fig:finitesize} demonstrate on one hand that despite reducing the three-dimensional magnetization vector field $\mathbf{m}$ to two coupled one-dimensional degrees of freedom—the wall position $u(x,t)$ and internal angle $\phi(x,t)$—the $u$--$\phi$ framework faithfully captures the essential physics of the driven interface as verified by full LLG simulations. Furthermore, we show that linear instabilities strongly affects the nonlinear dynamics in systems with a low density of unstable modes, in spite that the linearized dynamics of Eqs.~\eqref{eq:lineardynamics} is only valid for short times starting from the flat initial condition. This agreement establishes a rigorous foundation for addressing the more complex regime of many interacting unstable modes in thermodynamically large systems, where nonlinear interactions lead to fully developed spatiotemporal chaos.

\begin{figure}[t]
    \centering
    \includegraphics[width=\linewidth]{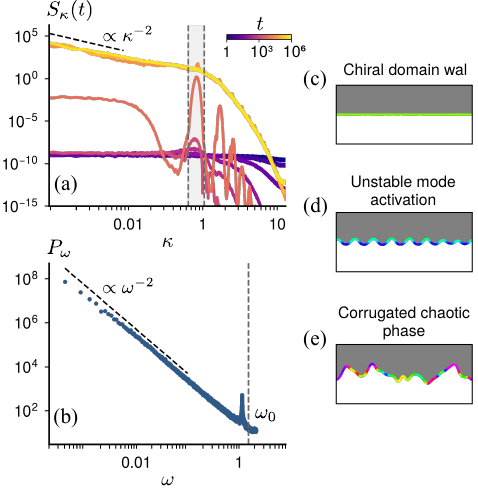}
    \caption{
    Transient dynamics from a flat, perturbed initial condition.
    (a) Time evolution of the $S_\kappa$ [Eq.~\eqref{eq:Skappa}] at \color{black}$t/\delta t=2^{0},2^{2},\dots,2^{18}$ with time step $\delta t=1.5$. \color{black} Shaded area marks the band of unstable modes at $h=3$.
    (b) Steady-state power spectrum [Eq.~\eqref{eq:powerspectrum}] of the DW center-of-mass velocity. Vertical line indicates  $\omega_0 \equiv 2\langle \dot \phi_1 \rangle$. Panels (c), (d) and (e) show DW snapshots at the initial condition, short and large times respectively (the color scale for $\phi$ corresponds to Fig.~\ref{fig:snap})
    }
    \label{fig:transient}
\end{figure}

\subsection{Spatiotemporal chaos and kinetic roughening}

\begin{figure*}[t]
    \centering
    \includegraphics[width=\linewidth]{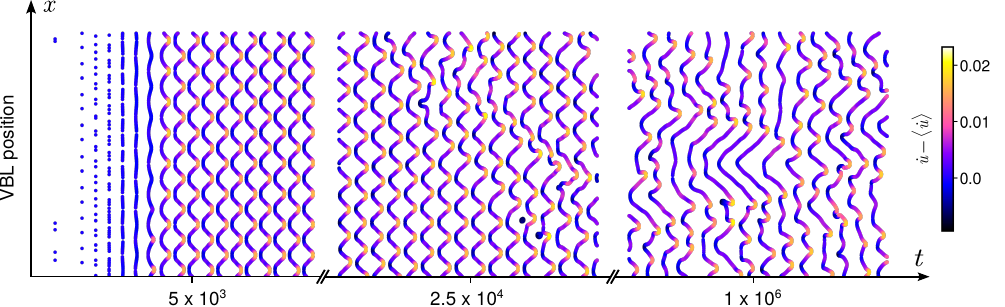}
    \caption{
        Snapshots of VBLs trajectories during the transient dynamics of Fig.~\ref{fig:transient}, at increasing times. Color indicates the DW velocity $\dot u(\xi,t)$ relative to $\langle \dot u \rangle$, at each VBL position $\xi$ (See Appendix~\ref{app:vbl}). 
    }    \label{fig:transient1}
\end{figure*}

We now consider large interfaces \(L/L_0 \gg 2\pi/\kappa_m\), for which many linearly unstable modes exist in the field range \(1<h<h_c\), and for each value of \(h\) within this interval at least one unstable mode is present [see Fig.~\ref{fig:stabilitydiag}(a)].
To quantify this behavior, Fig.~\ref{fig:heatmaps} in Appendix~\ref{sec:w2andvvshfinitesize} shows that already for \(L/\Delta \approx 300\) the deviations of the mean velocity \(v\) and roughness \(w^2\) from rigid-wall dynamics extend throughout the entire interval \(1<h<h_c\).
This contrasts with smaller systems, where finite-size effects give rise to stable windows within this range, leading to the discontinuous alternating features observed in Fig.~\ref{fig:finitesize}.

Large systems exhibit long transients and rich non-stationary dynamics, 
driven by the nonlinear $\sin(2 \phi)$ term in Eqs.~\eqref{eq:uphi}, which couples all modes and promotes a cascade of fluctuations across length scales. Figure~\ref{fig:transient}(a) shows the time evolution of the structure factor
\begin{equation}
S_\kappa(t) = \overline{ | \hat u_\kappa(t) |^2 }
\label{eq:Skappa}
\end{equation}
for \(L/\Delta = 16384\), starting from a weakly perturbed flat wall and where $\overline{(\ldots)}$ denote average over 200 realizations of the random initial condition. At early times, only stable high-\(\kappa\) modes decay with characteristic times \(\tau_\kappa \sim \kappa^{-2}\). Subsequently, unstable low-\(\kappa\) modes (shaded area Fig.~\ref{fig:transient}(a)) begin to grow, followed by spectral broadening due to nonlinear interactions. At late times, the system reaches a rough steady state with 
\(S_\kappa \sim \kappa^{-(1+2\zeta)}\), \(\zeta \approx 1/2\), and a center-of-mass velocity spectrum
\begin{equation}
P_\omega = \omega^2 \langle | \hat u_{\omega,\kappa=0} |^2 \rangle
\label{eq:powerspectrum}
\end{equation}
that exhibits broadband noise with \( \sim 1/\omega^2\) tails and a peak at the mean precession frequency near
\(\omega_0 = 2\langle \dot \phi_1 \rangle = \alpha  \sqrt{h^2-1}\) [Fig.~\ref{fig:transient}(b)]. In contrast to the perfectly periodic motion of rigid DWs and the nearly periodic corrugated phase in small systems, this large scale behavior indicates 
\color{black}
\textit{deterministic spatiotemporal chaos}
(see also LLG simulations in Appendix \ref{sec:emergenceofchaosinLLG}).

The non-steady relaxation is accompanied by complex VBL dynamics, shown in the space-time plots of Fig. \ref{fig:transient1} \footnote{See also movie S2 in the Supplemental Material \cite{supplemental}}. Initial isolated nucleation and annihilation events 
[Fig.~\ref{fig:transient1}(a)]
give way to near-periodic, collective trajectories [Fig.~\ref{fig:transient1}(b)]: VBL pairs nucleate, separate, and each annihilates with an oppositely moving partner. 
Interestingly, kink pair nucleation and annihilation typically occur at positions where the local DW velocity is, respectively, smaller or larger than the average velocity.
Eventually, these trajectories develop into a persistent intricate spatiotemporal  pattern [Fig.~\ref{fig:transient1}(c)]~\footnote{Ref.~\cite{skaugen2022} reports Bloch lines in large DWs even without disorder, attributing them to “numerical noise” and interpreting the effect as an “infinitesimal disorder case.” Our results indicate instead that these features reflect genuine spatiotemporal chaos arising from linearly unstable DW modes.}. 
This spatiotemporal chaos, mediated by the proliferation and turbulent dynamics of vertical Bloch lines, is reminiscent of the defect chaos observed in other extended nonlinear systems far from equilibrium~\cite{Aranson2002,cross2009pattern}.

\begin{figure}[t]
    \centering
\includegraphics[width=\linewidth]{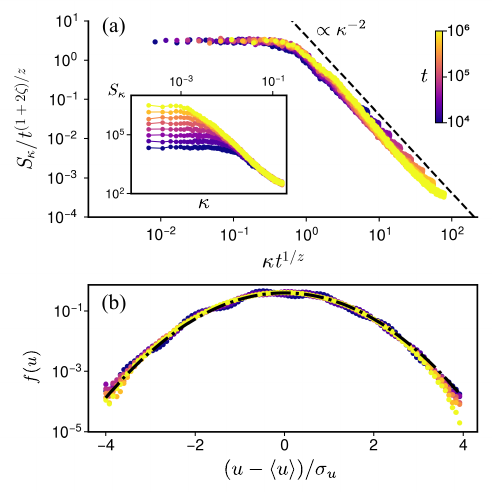}
    \caption{Kinetic roughening scaling of \( S_\kappa(t) \). (a) Data collapse onto a master curve using roughness exponent \( \zeta = 1/2 \) and dynamic exponent \( z = 2 \) \color{black}at times $t/\delta t=2^{15},2^{16},\dots,2^{22}$, \color{black} with $\delta t=4$. Dashed line shows \( \kappa^{-2} \) scaling. Inset: Raw \( S_\kappa \) data. (b) Distribution of DW height \( f(u) \), with \( \sigma_u^2 = \langle [u - \langle u \rangle]^2 \rangle \). Dash-dotted line is a Gaussian distribution.}
    \label{fig:universality}
\end{figure}

Identifying the universality class of the rough DW phase requires large systems and long times to reach a scaling regime. This process accelerates for values of \( h \) with large average growth rates \( |\mu_\kappa| \gg 1 \), which enhance nonlinear interactions.
Figure~\ref{fig:universality} presents results for \( h = 1.1 \) and \( L/\Delta = 131072 \) in the non-stationary regime. In Fig.~\ref{fig:universality}(a), the low-\( \kappa \) behavior of \( S_\kappa(t) \) follows kinetic roughening scaling~\cite{barabasi1995fractal}, 
\[
S_\kappa(t) \sim \kappa^{-(1+2\zeta)} G(\kappa t^{1/z}),
\]
with roughness exponent \( \zeta \approx 1/2 \) and dynamic exponent \( z \approx 2 \).
Fig.~\ref{fig:universality}(b) shows an approximately Gaussian height distribution \( f(u) \equiv \langle \delta(u(x,t)-u) \rangle \), consistent with the 1D Edwards--Wilkinson (EW) universality class. A slow crossover at larger scales to the 1D Kardar-Parisi-Zhang class 
(\( \zeta = 1/2, z = 3/2 \)), characterized by a non-Gaussian $f(u)$~\cite{takeuchi2011growing}, cannot be ruled out. Nonetheless, the length scales over which the 1D EW scaling consistently holds are experimentally relevant.

\section{Conclusions}

We have investigated the stability of dc-driven extended flat chiral domain walls and shown that, at zero temperature and in the absence of disorder, they undergo a dynamical phase transition at a critical field that we determine. This transition connects a flat chiral steady state to a corrugated, achiral, spatiotemporally chaotic nonequilibrium phase, characterized by universal roughening properties, linking domain wall dynamics with stochastic surface growth processes in general~\cite{barabasi1995fractal}. From a dynamical-systems perspective, our results thus provide a deterministic, disorder-free realization of spatiotemporal chaos arising solely from intrinsic nonlinear domain-wall dynamics. This behavior is reminiscent of that described by the Kuramoto–Sivashinsky equation and related nonlinear extended systems far from equilibrium~\cite{Cross1993, cross2009pattern,Aranson2002}.

Finite-size effects play a central role in the observed dynamics. The instability of the flat chiral domain wall emerges only above an experimentally relevant characteristic system size, which we identify. Below this threshold, the wall behaves as a rigid object following the standard Walker model for all fields. Above this scale, the system develops field-dependent dynamical instabilities, manifested as alternating windows of rigid and corrugated motion. These regimes are accurately predicted by a quasi-universal linear stability diagram controlled by the Gilbert damping. In sufficiently large systems, the rigid-wall windows disappear entirely above the Walker field, and the corrugated phase becomes fully chaotic. This chaotic regime persists up to a thermodynamic critical field, extending well beyond the region of negative differential mobility. Remarkably, this spatiotemporal chaos—characterized by achirality and complex Bloch-line dynamics (similar to the  ``defect chaos'' in other nonlinear extended systems \cite{Aranson2002})—terminates abruptly above the critical field, yielding a transition back to a stable, flat chiral moving domain wall.

Our results demonstrate that systems sharing the same Gilbert damping exhibit common instability mechanisms and dynamical phase transitions when fields, times, and lengths are expressed in appropriate nonuniversal units set by micromagnetic parameters. The framework introduced here can be naturally extended to include spin torque, Dzyaloshinskii–Moriya interaction, and higher-dimensional geometries, and provides a basis for addressing domain-wall dynamics in the presence of thermal fluctuations and quenched disorder by disentangling their effects from intrinsic deterministic instabilities.

\begin{acknowledgments}
We thank E. A. Jagla and T. Giamarchi for discussions. Supported by CNRS IRP “Statistical Physics of Materials”. Simulations used the Gerencia de Física  HPC cluster at CAB-CNEA (SNCAD, Argentina).
\end{acknowledgments}

\appendix

\section{$u$--$\phi$ model details}

\subsection{Derivation}
\label{App:uphi_deriv}

\begin{figure*}[t]
    \centering
    \includegraphics[width=\linewidth]{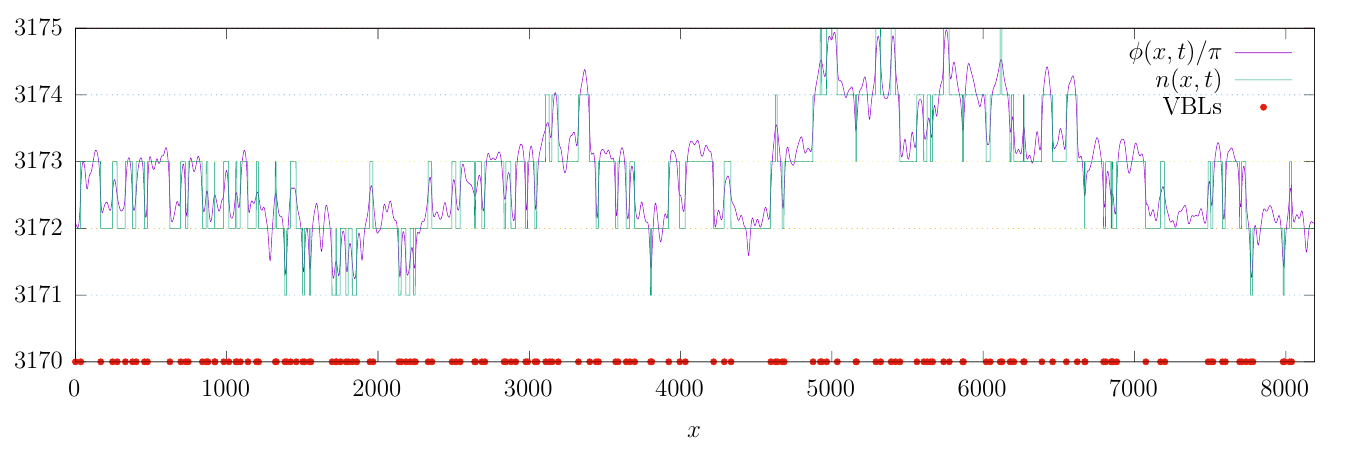}
    \caption{
        VBLs detection algorithm applied at a particular time $t$. The red dots indicate the detected VBLs positions, corresponding to the discontinuities of $n(x,t)$ [Eq.~\eqref{eq:nxt}].
    }
    \label{fig:vbl}
\end{figure*}

The equations of motion for the DW position $u(x,t)$ and magnetization angle $\phi(x,t)$ can be derived using Lagrangian formalism. The Lagrangian is given by

\begin{equation}
\mathcal{L} = \dfrac{\mu_{0}M_{s}}{\gamma}\int \dot \phi(1-\cos\theta)d\mathbf{x} - E
\end{equation}

where $E$ is the total energy defined in Eq.~\eqref{eq:energy}. Gilbert damping is incorporated through the Rayleigh dissipation function

\begin{equation}
    \mathcal{F} = \dfrac{\alpha \mu_{0}M_{s}}{2\gamma}\int(\dot{\theta}^{2}+\sin\theta\dot{\varphi}^{2})d\mathbf{x}
\end{equation}

Substituting the Bloch ansatz, Eqs.~\eqref{eq:Bloch2}, into these expressions and integrating over the $y$ yields one-dimensional functionals of $u(x,t)$ and $\phi(x,t)$. Under the assumption that the DW profile varies slowly in the $x$-direction (i.e., $|\partial u/\partial x|<<1$), the dynamics are governed by the Euler-Lagrange equations
\begin{equation} \dfrac{d}{dt} \left(\dfrac{\delta\mathcal{L}}{\delta \dot{X}}\right)-\dfrac{\delta\mathcal{L}}{\delta X}+ \dfrac{\partial \mathcal{F}}{\partial \dot{X}}=0, \end{equation}
where $X\in u,\phi$ and the functional derivative is defined as $\delta/\delta X=\partial/\partial X-\partial_{x} (\partial/\partial (\partial_{x}X) )$. Evaluating these equations yields (with prime denoting $\partial_{x}$):

\begin{align} \dot u &= \frac{\gamma M_s \Delta}{1+\alpha^2} \left[ \alpha \frac{H_z}{M_s} + \frac{N_n}{2} \sin 2\phi + \frac{2 A_\text{ex}}{\mu_0 M_s^2} \left( \frac{\alpha}{\Delta} u'' - \phi'' \right) \right],\\ \ \dot\phi &= \frac{\gamma M_s}{1+\alpha^2} \left[ \frac{H_z}{M_s} - \alpha \frac{N_n}{2} \sin 2\phi + \frac{2 A_\text{ex}}{\mu_0 M_s^2} \left( \frac{1}{\Delta} u'' + \alpha \phi'' \right) \right]. \end{align}

To extract the dimensionless form, we rescale the variables according to the natural scales of the problem: 

\begin{itemize}
    \item $x \to x/L_0$, with 
    \begin{align*}
            L_0 &= \Delta \sqrt{2\tilde K/N_n}, \\ \tilde{K} &= 2 K_u / (\mu_0 M_s^2) - 1.
    \end{align*}
    \item $u \to u/\Delta$,
    \item $t \to t/T_0$, with $T_0 = (1+\alpha^2)/(\gamma M_s N_n)$,
    \item $H_z \to h H_W$, with $H_W = \alpha M_s N_n/2$.
\end{itemize}

This non-isotropic scaling facilitates numerical implementation with different grid spacings in $x$ and $y$, only constrained by the smallest length scales. In these dimensionless units, the model becomes Eqs.~\eqref{eq:uphi} where all variables are dimensionless. Solutions apply to any system with the same $\alpha$ and $h$.

\subsection{Numerical Simulation}
\label{App:uphi_num}

We solve the nondimensionalized equations \eqref{eq:uphi} by following the approach of Ref.~\cite{skaugen2022}. Defining the complex field
\begin{equation}
    z(x,t) = u(x,t) - i\phi(x,t),
\end{equation}
the coupled real equations can be compactly written as a single complex evolution equation:
\begin{equation}
    \dot z = \frac{\alpha - i}{2} \left[ \alpha h + i \sin(2\,\mathrm{Im}(z)) + z'' \right].
\end{equation}

We use a pseudospectral Crank--Nicolson scheme to integrate this equation. While less efficient than real-space schemes for local dynamics, this method simplifies generalization to systems with nonlocal elasticity and higher dimensions,  and is convenient for analyzing the time evolution of spectral properties. We split the right-hand side into a linear and a nonlinear part:
\begin{align*}
    {\cal L}(z) &= \frac{\alpha - i}{2} z'', \\
    \mathcal{N}(z) &= \frac{\alpha - i}{2} \left[ \alpha h + i \sin(2\,\mathrm{Im}(z)) \right],
\end{align*}
and apply Crank--Nicolson to the linear term and an explicit Euler step to the nonlinear term:
\begin{equation}
    \frac{z^{n+1} - z^n}{\delta t}
    = \frac{1}{2} {\cal L}(z^{n+1} + z^n) + \mathcal{N}(z^n).
\end{equation}

Transforming to Fourier space, let \( \hat{z}_k^n \) denote the $k$-th Fourier mode of \( z^n(x) \). The linear operator becomes \( \hat{L}_k = -\frac{\alpha - i}{2} k^2 \), so the update rule for each mode reads:
\begin{equation}
    \left(1 - \frac{\delta t}{2} \hat{\cal L}_k\right) \hat{z}_k^{n+1}
    =
    \left(1 + \frac{\delta t}{2} \hat{\cal L}_k\right) \hat{z}_k^n + \delta t \, \widehat{\mathcal{N}(z^n)}_k.
\label{eq:cn_fourier}
\end{equation}

The algorithm proceeds as follows:
\begin{enumerate}
    \item Prepare the random initial condition $z(x,t=0)=r_u(x)-i r_\phi(x)$, with $r_u(x)$ and $r_\phi(x)$ independent uniformly distributed random numbers $\in [0,\epsilon)$ with $0<\epsilon \ll 1$. 
    \item Compute \( z^n(x) \) via inverse FFT.
    \item Extract \( \phi^n(x) = -\mathrm{Im}[z^n(x)] \).
    \item Evaluate \( \mathcal{N}(z^n) \) in real space.
    \item Transform \( \mathcal{N}(z^n) \to \widehat{\mathcal{N}(z^n)}_k \) via FFT.
    \item Update each mode \( \hat{z}_k^{n+1} \) using Eq.~\eqref{eq:cn_fourier}.
    \item Let $n+1 \to n$. Go to step 2.
\end{enumerate}

We choose the time step $\delta t$ small enough to ensure convergence and independence of the results from $\delta t$. As a validation check, we confirm that the dynamics of a rigid wall (neglecting $z''$) match the analytical solution. We also verify that the onset of instabilities remains unchanged with varying $\delta t$. The spatial resolution is chosen to resolve the narrowest physical feature of the system, the internal structure of a kink. In practice, we impose a spatial step size well below the kink width to ensure numerical accuracy. The simulation code was written in \texttt{CUDA C/C++}, using \texttt{cuFFT}, \texttt{Thrust}, and custom \texttt{CUDA} kernels~\cite{cuda2024}.

\subsection{Vertical Bloch Lines}
\label{app:vbl}

When the DW distortion is moderate (no overhangs), vertical Bloch lines (VBLs) can be identified as kinks in the field 
$\phi(x,t)$, corresponding to localized, smooth changes of approximately $\pi$. To detect and precisely locate these VBLs, we first define a discrete integer-valued field $n(x,t) \in \mathbb{Z}$ as

\begin{equation}
    n(x,t) \equiv \left\lfloor \frac{\phi(x,t)}{\pi} + \frac{1}{2} \right\rfloor,
    \label{eq:nxt}
\end{equation}

where \(\lfloor \cdot \rfloor\) denotes the integer part (floor function).

A vertical Bloch line is identified whenever there is a discontinuity such that 
\begin{equation}
|n(x^*,t) - n(x^*+1,t)| = 1,
\end{equation}
and its position $\xi$ at time \(t\) is assigned to \(x^* + \frac{1}{2}\) (See Fig.~\ref{fig:transient1} in the main text).

In Figure~\ref{fig:vbl}, we illustrate the detection procedure, which can be efficiently implemented for large interfaces using parallel algorithms. See Movie~S2 in the Movies section to watch this algorithm in action.

\section{Matching length scales between LLG and $u$–$\phi$ model}
\label{app:matchingscales}

\begin{figure}[h]
    \centering
\includegraphics[]{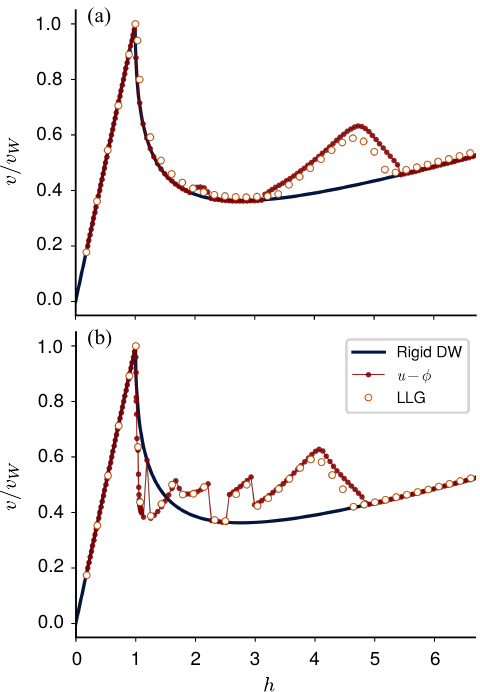}
    \caption{
    Velocity field characteristics for moving DWs obtained from simulations of the $u$–$\phi$ model (system size $L$) and corresponding LLG simulations (system size $L' = L \Delta' / \Delta$), using the same micromagnetic parameters: $N_n = 0.01667$, $\alpha = 0.27$, $A_{ex} = 1.4 \times 10^{-11}$~J/m, $M_s = 9.1 \times 10^5$~A/m, and $K_u = 8.4 \times 10^5$~J/m³, as in Ref.~\cite{skaugen2022}.
    (a) $L/\Delta = L'/\Delta' = 51$. (b) $L/\Delta = L'/\Delta' = 112$.
    }
    \label{fig:matching}
\end{figure}
The ansatz underlying the $u$--$\phi$ model relies on several assumptions. It assumes that the DW can be described as a single-valued function of $x$, that the azimuthal angle of $\mathbf{m}$ is independent of $y$, and that the DW profile along $y$ is indistinguishable from its equilibrium form. In particular, it neglects DW width fluctuations, fixing the width to the value
\[
\Delta = \sqrt{\frac{A_\text{ex}}{K_0}} .
\]
This choice sets the unit $L_0$ used to measure distances in $x$ (see previous Section), which must be related to the scale used in LLG simulations or experiments.  
Furthermore, the ansatz is introduced within an already simplified micromagnetic model, especially regarding the magnetostatic contribution~\cite{skaugen2019}.

When comparing with LLG simulations, we have empirically found that the best agreement with the $u$--$\phi$ modes is obtained by applying a correction factor of order $\Delta/\Delta'$, where $\Delta' \approx 6.356$ nm is the average DW width measured in the LLG simulations, and $\Delta \approx 6.62$ nm for the parameters of Fig.~\ref{fig:LLG1}. This small difference between $\Delta$ and $\Delta'$ likely arises from dynamical effects and the residual influence of the full magnetostatic interaction, which are not captured in the simplified $u$–$\phi$ model. Therefore, to match a system of size $L$ in the $u$--$\phi$ model, we choose the corresponding size in the LLG simulations as $L' = L (\Delta' / \Delta)$. Although $\Delta/\Delta' \approx 1.041$ is close to unity, we have consistently found that this correction is necessary to achieve optimal agreement in both the velocity and roughness field characteristics.

The proposed matching is tested and illustrated in Fig.~\ref{fig:matching}, where we compare the velocity field characteristics for two $u$–$\phi$ simulations with $L/\Delta = 112$ and $L/\Delta = 51$ against \texttt{Mumax$^3$} simulations using the corresponding $L'$. We observe good overall agreement, except in field ranges where the linear instabilities are weak (i.e., where the absolute value of the Floquet multipliers is only slightly above unity, the darker regions in Fig.~\ref{fig:gallery}).

\section{Linear Stability}

\subsection{Gallery of linear stability maps}
\label{app:LS}

\begin{figure*}[t]
    \centering
    \includegraphics[width=\linewidth]{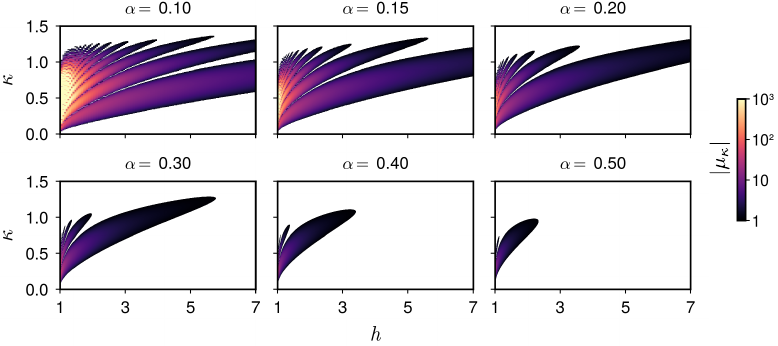}
    \caption{Linear stability diagrams showing $|\mu_{\kappa}|$ as a function of wavenumber $\kappa$
 and driving field $h$ for different values of the damping parameter $\alpha$. Colored regions indicate unstable modes ($|\mu_{\kappa}|>1$).}
    \label{fig:gallery}
\end{figure*}

Fig.~\ref{fig:gallery} shows spectral stability maps as a function of the Gilbert damping parameter $\alpha$ as obtained from the stability analysis of flat DWs in the $u$--$\phi$ model. For each field value $h$, the maps indicate which DW mode wavevectors become linearly unstable, and the color encodes their growth rate. Watch also the movie S1~\cite{supplemental}.

\subsection{Below the Walker Field}
\label{sec:belowwalkerfield}
For $h < 1$, we have $\dot{\phi}_1 = 0$ and $\phi_1 = \arcsin(h)/2$.  
In this regime, the eigenvalues of the dynamical matrix in Eq.~\eqref{eq:linearabove} are  
\begin{align*}
\lambda_{\pm} &= -\frac{\alpha}{2} \left( \sqrt{1-h^2} + \kappa^2 \right)
\\ &\pm \frac{1}{2}\sqrt{\alpha^2(1-h^2) - \kappa^4 - 2\kappa^2\sqrt{1-h^2}}.
\end{align*}

Stability is more easily assessed via the trace and determinant of Eq.~\eqref{eq:linearabove}. The trace is  
\begin{equation}
\operatorname{Tr} = -\frac{\alpha}{2} \left( \sqrt{1-h^2} + \kappa^2 \right) < 0,
\end{equation}
while the determinant is  
\begin{align}
    \det &= \frac{\alpha \kappa^2}{2}\left( \frac{\alpha \kappa^2}{2} + \alpha \sqrt{1-h^2} \right) \nonumber
\\&+ \frac{\kappa^2}{2} \left( \frac{\kappa^2}{2} + \sqrt{1-h^2} \right) > 0,
\end{align}
for $\kappa^2 > 0$.  

Thus, the system has either a stable node or a stable spiral~\cite{strogatz2015}, with both eigenvalues having negative real parts. The general solution is a linear combination of decaying exponentials, $e^{\lambda_{\pm} t}$, so the flat DW (uniform in $u$ and $\phi$) is stable in the stationary regime below the precessional threshold. In other words, for $h<1$,  
\begin{equation}
u(x,t) = u_1(t), \quad \phi(x,t) = \frac{\arcsin(h)}{2}.
\end{equation}

\subsection{Roughness and mean velocity as a function of the system size}
\label{sec:w2andvvshfinitesize}

\begin{figure}[h]
    \centering
    \includegraphics[width=\linewidth]{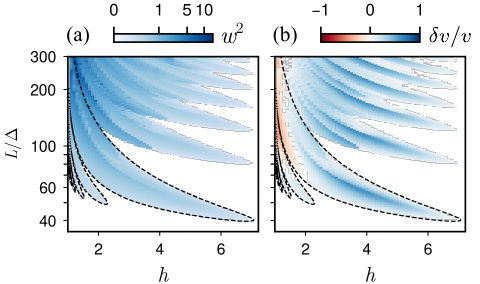}
    \caption{Heat maps of the steady-state roughness (a) and relative velocity $\delta v/v$ computed using the $u$--$\phi$ model.}
    \label{fig:heatmaps}
\end{figure}

Figures~\ref{fig:heatmaps}(a) and \ref{fig:heatmaps}(b) show heat maps of $w^2$ and the relative mean velocity difference $\delta v / v$ via

\begin{align}
{\delta v}/{v} = {\langle \partial_t u \rangle}/{\langle \dot u_1 \rangle}-1, 
\label{eq:relativevel}
\end{align}
from $u$--$\phi$ simulations with $\alpha = 0.27$, over the $(L/\Delta, h)$ plane.

Corrugation sets in near $L = L_0(2\pi/\kappa_m)$ and $h \approx h_c$, expanding into a broad band in $h$ and followed by narrow downward-extending “feathers”. Strikingly, the dashed lines match the linear stability threshold from Fig.~\ref{fig:stabilitydiag}(a), re-parametrized in terms of $2\pi/\kappa$, confirming that fundamental mode instability drives the onset. As $L$ increases, additional feather-like features emerge, tracing instabilities of higher harmonics falling into the feathers of Fig.~\ref{fig:stabilitydiag}, while small systems exhibit simpler behavior due to sparse mode availability.

\section{Emergence of spatiotemporal chaos in LLG simulations}
\label{sec:emergenceofchaosinLLG}

As shown in the main text with the $u$--$\phi$ model, long interfaces ($L \gg \Delta$) can excite many modes simultaneously, and nonlinear interactions lead to spatiotemporal chaos with spectral broadening. These results are confirmed by LLG simulations using \texttt{Mumax³} \cite{vansteenkiste2014design}, with \texttt{Ubergmag} \cite{beg2022} and \texttt{Oommfpy}~\cite{cortes2019} employed for initial state preparation and data processing.

We consider a thin film with the micromagnetic parameters detailed in Section~\ref{sect:mmag}, with the following geometry: $L_x = 4096$ nm $\approx 619\Delta$, $L_y = 2048$ nm, $L_z = 0.5$ nm. The applied magnetic field is $\mu_{0}H_{z}=6$ mT ($h\simeq 2.8$) for a total simulation time $T=400$ ns, with temporal sampling $\Delta t=0.2$ ns. If $u(x,0)$ is constant, the DW remains flat during evolution, and instabilities cannot be tested. Therefore, we use
\begin{equation}
     u(x,0) =A\sum_{n=1000}^{1023}\sin\left(\frac{2\pi n}{L}x+\beta_{n}\right)
\end{equation}
where $\beta_{n}$ is uniformly distributed in $[0,2\pi)$ and $A=0.01$ nm. We put $\phi(x,t)=0$. The exact form of the random initial perturbation is not important.

\begin{figure}[t]
    \centering
    \includegraphics[width=\linewidth]{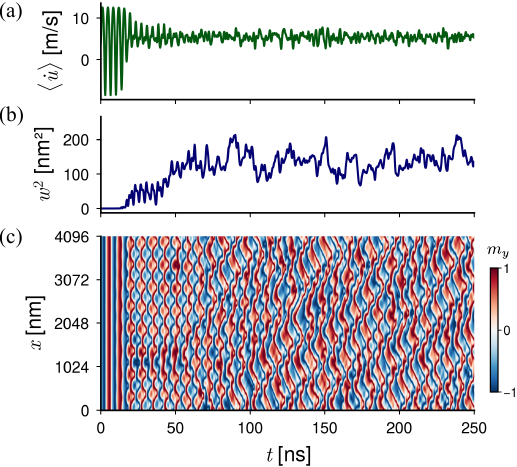}
    \caption{Transient dynamics for a Co/Pt thin film with $L=4096$ nm at $h=2.8$ obtained from LLG simulations. Temporal evolution of the average wall velocity $\langle \dot{u} \rangle$ (a), roughness $w^{2}$ (b), and spatiotemporal map of $m_{y}$ over the wall (c).}
    \label{fig:LLG1}
\end{figure}

The temporal evolution of the center-of-mass velocity and the roughness are shown in Fig.~\ref{fig:LLG1}(a) and Fig.~\ref{fig:LLG1}(b), respectively, while Fig.~\ref{fig:LLG1}(c) presents a space–time map of the $m_{y}$ component along the wall. At short times, magnetization precession is observed with $w^2 \approx 0$, corresponding to dynamics similar to that of a rigid wall. Subsequently, unstable modes lead to the nucleation of pairs of vertical Bloch lines ($m_{y} \approx \pm 1$) that travel along the wall until annihilation, following a nearly periodic pattern in $x$. After approximately $t \approx 50$ ns, the roughness increases and the nucleation–annihilation sites lose their spatial regularity (see Movie S3, Fig.~\ref{fig:vbl} in the main text, and Movie S2 for similar results in the $u$--$\phi$).

\begin{figure}[t]
    \centering
    \includegraphics[width=\linewidth]{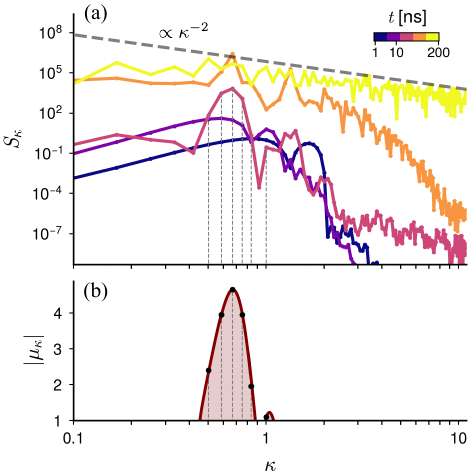}
    \caption{Temporal evolution of the structure factor obtained from LLG simulations (a) and the absolute value of the Floquet multipliers $\mu_{\kappa}$ at field $h=2.8$ obtained by solving the linearized $u$--$\phi$ model (b). Dashed vertical lines indicate unstable modes.}
    \label{fig:LLG2}
\end{figure}

Fig. \ref{fig:LLG2}(a) shows the structure factor at different times for the system analyzed in Fig.~\ref{fig:LLG1}. Analogously to Fig.~\ref{fig:transient} of the main text, there is initially a decay of high-$\kappa$ modes. As time progresses, unstable modes appear whose Floquet multipliers satisfy $\lvert \mu_{\kappa} \rvert > 1$ [see Fig.~\ref{fig:LLG2}(b)]. Finally, at long times a stationary regime is reached in which $S_{\kappa} \sim \kappa^{-2}$ over increasingly large length scales, consistent with the kinetic roughening scaling of Fig.~\ref{fig:universality} in the main text.

\bibliography{refs}   

@article{Schryer1974,
  author    = {N. L. Schryer and L. R. Walker},
  title     = {The motion of 180\textdegree{} domain walls in uniform dc magnetic fields},
  journal   = {Journal of Applied Physics},
  volume    = {45},
  number    = {12},
  pages     = {5406--5421},
  year      = {1974},
  doi       = {10.1063/1.1663252},
  url       = {https://doi.org/10.1063/1.1663252},
  publisher = {American Institute of Physics}
}

@inproceedings{Slonczewski1972,
  author    = {J. C. Slonczewski},
  title     = {Dynamics of Magnetic Domain Walls},
  booktitle = {AIP Conference Proceedings},
  volume    = {5},
  number    = {1},
  pages     = {170--174},
  year      = {1972},
  doi       = {10.1063/1.2946580},
  url       = {https://doi.org/10.1063/1.2946580},
  publisher = {American Institute of Physics}
}

@article{skaugen2022,
  title = {Depinning Exponents of Thin Film Domain Walls Depend on Disorder Strength},
  author = {Skaugen, Audun and Laurson, Lasse},
  journal = {Phys. Rev. Lett.},
  volume = {128},
  issue = {9},
  pages = {097202},
  numpages = {5},
  year = {2022},
  month = {Mar},
  publisher = {American Physical Society},
  doi = {10.1103/PhysRevLett.128.097202},
  url = {https://link.aps.org/doi/10.1103/PhysRevLett.128.097202}
}

@article{thiaville2005,
doi = {10.1209/epl/i2004-10452-6},
url = {https://dx.doi.org/10.1209/epl/i2004-10452-6},
year = {2005},
month = {feb},
publisher = {},
volume = {69},
number = {6},
pages = {990},
author = {A. Thiaville and Y. Nakatani and J. Miltat and Y. Suzuki},
title = {Micromagnetic understanding of current-driven 
domain wall motion in patterned nanowires},
journal = {Europhysics Letters}
}

@book{malozemoff1979magnetic,
  title     = {Magnetic Domain Walls in Bubble Materials},
  author    = {Malozemoff, A.P. and Slonczewski, J.C.},
  year      = {1979},
  publisher = {Academic Press},
  address   = {New York},
  isbn      = {9780120029502}
}

@article{dipietro2023,
  title = {Domain wall statics and dynamics in nanowires with arbitrary Dzyaloshinskii-Moriya tensors},
  author = {Di Pietro, Adriano and Garc\'{\i}a-S\'anchez, Felipe and Durin, Gianfranco},
  journal = {Phys. Rev. B},
  volume = {108},
  issue = {17},
  pages = {174427},
  numpages = {10},
  year = {2023},
  month = {Nov},
  publisher = {American Physical Society},
  doi = {10.1103/PhysRevB.108.174427},
  url = {https://link.aps.org/doi/10.1103/PhysRevB.108.174427}
}

@article{Parkin2008,
  title = {Magnetic Domain-Wall Racetrack Memory},
  author = {Parkin, Stuart S. P. and Hayashi, Masamitsu and Thomas, Luc},
  journal = {Science},
  volume = {320},
  number = {5873},
  pages = {190--194},
  year = {2008},
  doi = {10.1126/science.1145799},
  url = {https://www.science.org/doi/10.1126/science.1145799}
}

@article{lemerle1998domain,
  title = {Domain Wall Creep in an Ising Ultrathin Magnetic Film},
  author = {Lemerle, S. and Ferr\'e, J. and Chappert, C. and Mathet, V. and Giamarchi, T. and Le Doussal, P.},
  journal = {Phys. Rev. Lett.},
  volume = {80},
  issue = {4},
  pages = {849--852},
  numpages = {0},
  year = {1998},
  month = {Jan},
  publisher = {American Physical Society},
  doi = {10.1103/PhysRevLett.80.849},
  url = {https://link.aps.org/doi/10.1103/PhysRevLett.80.849}
}

@article{chauve2000creep,
  title={Creep and depinning in disordered media},
  author={Chauve, Pascal and Giamarchi, Thierry and Le Doussal, Pierre},
  journal={Physical Review B},
  volume={62},
  number={10},
  pages={6241},
  year={2000},
  publisher={APS}
}

@article{ferrero2021creep,
  title={Creep motion of elastic interfaces driven in a disordered landscape},
  author={Ferrero, Ezequiel E and Foini, Laura and Giamarchi, Thierry and Kolton, Alejandro B and Rosso, Alberto},
  journal={Annual Review of Condensed Matter Physics},
  volume={12},
  number={1},
  pages={111--134},
  year={2021},
  publisher={Annual Reviews}
}

@article{beach2005dynamics,
  title={Dynamics of field-driven domain-wall propagation in ferromagnetic nanowires},
  author={Beach, Geoffrey S D and Nistor, Claudiu and Knutson, Curtis and Tsoi, Maxim and Erskine, James L},
  journal={Nature Materials},
  volume={4},
  number={10},
  pages={741--744},
  year={2005},
  publisher={Nature Publishing Group},
  doi={10.1038/nmat1477}
}

@article{Dourlat2008,
  title = {Field-driven domain-wall dynamics in (Ga,Mn)As films with perpendicular anisotropy},
  author = {Dourlat, A. and Jeudy, V. and Lema\^{\i}tre, A. and Gourdon, C.},
  journal = {Phys. Rev. B},
  volume = {78},
  issue = {16},
  pages = {161303},
  numpages = {4},
  year = {2008},
  month = {Oct},
  publisher = {American Physical Society},
  doi = {10.1103/PhysRevB.78.161303},
  url = {https://link.aps.org/doi/10.1103/PhysRevB.78.161303}
}

@article{Jeudy2016,
  title = {Universal Pinning Energy Barrier for Driven Domain Walls in Thin Ferromagnetic Films},
  author = {Jeudy, V. and Mougin, A. and Bustingorry, S. and Savero Torres, W. and Gorchon, J. and Kolton, A. B. and Lema\^{\i}tre, A. and Jamet, J.-P.},
  journal = {Phys. Rev. Lett.},
  volume = {117},
  issue = {5},
  pages = {057201},
  numpages = {5},
  year = {2016},
  month = {Jul},
  publisher = {American Physical Society},
  doi = {10.1103/PhysRevLett.117.057201},
  url = {https://link.aps.org/doi/10.1103/PhysRevLett.117.057201}
}

@article{Albornoz2021,
  title = {Universal critical exponents of the magnetic domain wall depinning transition},
  author = {Albornoz, L. J. and Ferrero, E. E. and Kolton, A. B. and Jeudy, V. and Bustingorry, S. and Curiale, J.},
  journal = {Phys. Rev. B},
  volume = {104},
  issue = {6},
  pages = {L060404},
  numpages = {5},
  year = {2021},
  month = {Aug},
  publisher = {American Physical Society},
  doi = {10.1103/PhysRevB.104.L060404},
  url = {https://link.aps.org/doi/10.1103/PhysRevB.104.L060404}
}

@article{Grassi2018,
  title = {Intermittent collective dynamics of domain walls in the creep regime},
  author = {Grassi, Mat\'{\i}as Pablo and Kolton, Alejandro B. and Jeudy, Vincent and Mougin, Alexandra and Bustingorry, Sebastian and Curiale, Javier},
  journal = {Phys. Rev. B},
  volume = {98},
  issue = {22},
  pages = {224201},
  numpages = {10},
  year = {2018},
  month = {Dec},
  publisher = {American Physical Society},
  doi = {10.1103/PhysRevB.98.224201},
  url = {https://link.aps.org/doi/10.1103/PhysRevB.98.224201}
}

@article{Pardo2017,
  title = {Universal depinning transition of domain walls in ultrathin ferromagnets},
  author = {Diaz Pardo, R. and Savero Torres, W. and Kolton, A. B. and Bustingorry, S. and Jeudy, V.},
  journal = {Phys. Rev. B},
  volume = {95},
  issue = {18},
  pages = {184434},
  numpages = {7},
  year = {2017},
  month = {May},
  publisher = {American Physical Society},
  doi = {10.1103/PhysRevB.95.184434},
  url = {https://link.aps.org/doi/10.1103/PhysRevB.95.184434}
}

@article{Durin2024,
  title = {Earthquakelike dynamics in ultrathin magnetic films},
  author = {Durin, Gianfranco and Schimmenti, Vincenzo Maria and Baiesi, Marco and Casiraghi, Arianna and Magni, Alessandro and Herrera-Diez, Liza and Ravelosona, Dafin\'e and Foini, Laura and Rosso, Alberto},
  journal = {Phys. Rev. B},
  volume = {110},
  issue = {2},
  pages = {L020405},
  numpages = {5},
  year = {2024},
  month = {Jul},
  publisher = {American Physical Society},
  doi = {10.1103/PhysRevB.110.L020405},
  url = {https://link.aps.org/doi/10.1103/PhysRevB.110.L020405}
}

@article{Hutner2019,
  title = {Multistep Bloch-line-mediated Walker breakdown in ferromagnetic strips},
  author = {H\"utner, Johanna and Herranen, Touko and Laurson, Lasse},
  journal = {Phys. Rev. B},
  volume = {99},
  issue = {17},
  pages = {174427},
  numpages = {5},
  year = {2019},
  month = {May},
  publisher = {American Physical Society},
  doi = {10.1103/PhysRevB.99.174427},
  url = {https://link.aps.org/doi/10.1103/PhysRevB.99.174427}
}

@article{gourdon2013domain,
  title = {Domain-wall flexing instability and propagation in thin ferromagnetic films},
  author = {Gourdon, C. and Thevenard, L. and Haghgoo, S. and C\ifmmode \bar{e}\else \={e}\fi{}bers, A.},
  journal = {Phys. Rev. B},
  volume = {88},
  issue = {1},
  pages = {014428},
  numpages = {9},
  year = {2013},
  month = {Jul},
  publisher = {American Physical Society},
  doi = {10.1103/PhysRevB.88.014428},
  url = {https://link.aps.org/doi/10.1103/PhysRevB.88.014428}
}

@article{thevenard2011domain,
  title = {Domain wall propagation in ferromagnetic semiconductors: Beyond the one-dimensional model},
  author = {Thevenard, L. and Gourdon, C. and Haghgoo, S. and Adam, J-P. and von Bardeleben, H. J. and Lema\^{\i}tre, A. and Schoch, W. and Thiaville, A.},
  journal = {Phys. Rev. B},
  volume = {83},
  issue = {24},
  pages = {245211},
  numpages = {9},
  year = {2011},
  month = {Jun},
  publisher = {American Physical Society},
  doi = {10.1103/PhysRevB.83.245211},
  url = {https://link.aps.org/doi/10.1103/PhysRevB.83.245211}
}

@book{hillebrands2006spin,
  title     = {Spin Dynamics in Confined Magnetic Structures III},
  editor    = {Hillebrands, Burkard and Thiaville, Andr{\'e}},
  series    = {Topics in Applied Physics},
  volume    = {101},
  publisher = {Springer Berlin Heidelberg},
  address   = {Berlin, Heidelberg},
  year      = {2006},
  isbn      = {978-3-540-39842-4},
  doi       = {10.1007/b12462},
}

@article{vansteenkiste2014design,
  title     = {The design and verification of MuMax3},
  author    = {Vansteenkiste, Arne and Leliaert, Jonathan and Dvornik, Mykola and Helsen, Mathias and Garcia-Sanchez, Felipe and Van Waeyenberge, Bartel},
  journal   = {AIP Advances},
  volume    = {4},
  number    = {10},
  pages     = {107133},
  year      = {2014},
  publisher = {AIP Publishing},
  doi       = {10.1063/1.4899186},
  url       = {https://doi.org/10.1063/1.4899186}
}

@article{krizakova2019,
  title = {Study of the velocity plateau of Dzyaloshinskii domain walls},
  author = {Krizakova, V. and {Pe\~na Garc\'{\i}a}, J. and Vogel, J. and Rougemaille, N. and {de Souza Chaves}, D.  and Pizzini, S. and Thiaville, A.},
  journal = {Phys. Rev. B},
  volume = {100},
  issue = {21},
  pages = {214404},
  numpages = {9},
  year = {2019},
  month = {Dec},
  publisher = {American Physical Society},
  doi = {10.1103/PhysRevB.100.214404},
  url = {https://link.aps.org/doi/10.1103/PhysRevB.100.214404}
}

@article{Cross1993,
  title = {Pattern formation outside of equilibrium},
  author = {Cross, M. C. and Hohenberg, P. C.},
  journal = {Rev. Mod. Phys.},
  volume = {65},
  issue = {3},
  pages = {851--1112},
  numpages = {0},
  year = {1993},
  month = {Jul},
  publisher = {American Physical Society},
  doi = {10.1103/RevModPhys.65.851},
  url = {https://link.aps.org/doi/10.1103/RevModPhys.65.851}
}

@misc{supplemental,
  title = {Supplemental Material},
  note  = {See Supplemental Material at [URL or DOI] for technical details.},
}

@book{barabasi1995fractal,
  title={Fractal Concepts in Surface Growth},
  author={Barab{\'a}si, Albert-L{\'a}szl{\'o} and Stanley, H.~Eugene},
  year={1995},
  publisher={Cambridge University Press},
  address={Cambridge},
  isbn={9780521483186}
}

@article{takeuchi2011growing,
  title={Growing interfaces uncover universal fluctuations behind scale invariance},
  author={Takeuchi, Kazumasa A. and Sano, Masaki and Sasamoto, Tomohiro and Spohn, Herbert},
  journal={Scientific Reports},
  volume={1},
  pages={34},
  year={2011},
  publisher={Nature Publishing Group},
  doi={10.1038/srep00034}
}

@book{cross2009pattern,
  title={Pattern Formation and Dynamics in Nonequilibrium Systems},
  author={Cross, M. C. and Greenside, H. S.},
  year={2009},
  publisher={Cambridge University Press},
  address={Cambridge},
  isbn={9780521770502}
}

@article{allwood2005magnetic,
  title = {Magnetic domain-wall logic},
  author = {Allwood, D. A. and Xiong, G. and Faulkner, C. C. and Atkinson, D. and Petit, D. and Cowburn, R. P.},
  journal = {Science},
  volume = {309},
  number = {5741},
  pages = {1688--1692},
  year = {2005},
  doi = {10.1126/science.1108813},
  publisher = {American Association for the Advancement of Science}
}

@article{Venkat_2024,
doi = {10.1088/1361-6463/ad0568},
url = {https://dx.doi.org/10.1088/1361-6463/ad0568},
year = {2023},
month = {nov},
publisher = {IOP Publishing},
volume = {57},
number = {6},
pages = {063001},
author = {Venkat, G and Allwood, D A and Hayward, T J},
title = {Magnetic domain walls: types, processes and applications},
journal = {Journal of Physics D: Applied Physics}
}

@article{kumar2022domain,
  title={Domain wall memory: Physics, materials, and devices},
  author={Kumar, Durgesh and Jin, Tianli and Sbiaa, Rachid and Kl{\"a}ui, Mathias and Bedanta, Subhankar and Fukami, Shunsuke and Ravelosona, Dafine and Yang, See-Hun and Liu, Xiaoxi and Piramanayagam, SN},
  journal={Physics Reports},
  volume={958},
  pages={1--35},
  year={2022},
  publisher={Elsevier}
}

@article{khovenkil2003corrugation,
  title = {Corrugation‐type instability of the free motion of 180° domain walls in uniaxial ferromagnets},
  author = {Khodenkov, G.~E.},
  journal = {Technical Physics Letters},
  volume = {29},
  number = {11},
  pages = {907--909},
  year = {2003},
  doi = {10.1134/1.1631360}
}

@book{hubert1998magnetic,
  title     = {Magnetic Domains: The Analysis of Magnetic Microstructures},
  author    = {Hubert, Alex and Schäfer, Rudolf},
  publisher = {Springer},
  address   = {Berlin},
  year      = {1998},
  isbn      = {978-3-540-64108-7},
  doi       = {10.1007/978-3-540-85054-0}
}

@article{Kim2024,
  title = {Enhancement of magnetic domain wall velocity via resonant dissipation of standing wave modes of domain wall structure with perpendicular magnetic anisotropy},
  author = {Kim, Ganghwi and Jung, Dae-Han and Han, Hee-Sung and Kang, Myeonghwan and Jeong, Suyeong and Park, Younggun and Im, Mi-Young and Lee, Ki-Suk},
  journal = {Phys. Rev. B},
  volume = {110},
  issue = {21},
  pages = {214407},
  numpages = {10},
  year = {2024},
  month = {Dec},
  publisher = {American Physical Society},
  doi = {10.1103/PhysRevB.110.214407},
  url = {https://link.aps.org/doi/10.1103/PhysRevB.110.214407}
}

@article{kolton2023,
  title = {Depinning free of the elastic approximation},
  author = {Kolton, A. B. and Ferrero, E. E. and Rosso, A.},
  journal = {Phys. Rev. B},
  volume = {108},
  issue = {17},
  pages = {174201},
  numpages = {8},
  year = {2023},
  month = {Nov},
  publisher = {American Physical Society},
  doi = {10.1103/PhysRevB.108.174201},
  url = {https://link.aps.org/doi/10.1103/PhysRevB.108.174201}
}

@article{skaugen2019,
  title = {Analytical computation of the demagnetizing energy of thin-film domain walls},
  author = {Skaugen, Audun and Murray, Peyton and Laurson, Lasse},
  journal = {Phys. Rev. B},
  volume = {100},
  issue = {9},
  pages = {094440},
  numpages = {10},
  year = {2019},
  month = {Sep},
  publisher = {American Physical Society},
  doi = {10.1103/PhysRevB.100.094440},
  url = {https://link.aps.org/doi/10.1103/PhysRevB.100.094440}
}

@article{Mougin2007,
  author = {Mougin, Alexandra and Cormier, Morgan and Adam, Jean-Paul and Metaxas, Peter J. and Ferré, J.},
  title = {Domain wall mobility, stability and Walker breakdown in magnetic nanowires},
  journal = {Europhysics Letters (EPL)},
  volume = {78},
  number = {5},
  pages = {57007},
  year = {2007},
  doi = {10.1209/0295-5075/78/57007}
}

@article{Garcia2021,
  title = {Magnetic domain wall dynamics in the precessional regime: Influence of the Dzyaloshinskii-Moriya interaction},
  author = {Garcia, Jose Pe\~na and Fassatoui, Aymen and Bonfim, Marlio and Vogel, Jan and Thiaville, Andr\'e and Pizzini, Stefania},
  journal = {Phys. Rev. B},
  volume = {104},
  issue = {1},
  pages = {014405},
  numpages = {9},
  year = {2021},
  month = {Jul},
  publisher = {American Physical Society},
  doi = {10.1103/PhysRevB.104.014405},
  url = {https://link.aps.org/doi/10.1103/PhysRevB.104.014405}
}

@article{Aranson2002,
  title = {The world of the complex Ginzburg-Landau equation},
  author = {Aranson, Igor S. and Kramer, Lorenz},
  journal = {Rev. Mod. Phys.},
  volume = {74},
  issue = {1},
  pages = {99--143},
  numpages = {0},
  year = {2002},
  month = {Feb},
  publisher = {American Physical Society},
  doi = {10.1103/RevModPhys.74.99},
  url = {https://link.aps.org/doi/10.1103/RevModPhys.74.99}
}

@article{Metaxas2007,
  title = {Creep and Flow Regimes of Magnetic Domain-Wall Motion in Ultrathin $\mathrm{Pt}/\mathrm{Co}/\mathrm{Pt}$ Films with Perpendicular Anisotropy},
  author = {Metaxas, P. J. and Jamet, J. P. and Mougin, A. and Cormier, M. and Ferr\'e, J. and Baltz, V. and Rodmacq, B. and Dieny, B. and Stamps, R. L.},
  journal = {Phys. Rev. Lett.},
  volume = {99},
  issue = {21},
  pages = {217208},
  numpages = {4},
  year = {2007},
  month = {Nov},
  publisher = {American Physical Society},
  doi = {10.1103/PhysRevLett.99.217208},
  url = {https://link.aps.org/doi/10.1103/PhysRevLett.99.217208}
}

@book{Strogatz2015,
  author    = {Steven H. Strogatz},
  title     = {Nonlinear Dynamics and Chaos: With Applications to Physics, Biology, Chemistry, and Engineering},
  edition   = {2},
  publisher = {Westview Press},
  year      = {2015},
  isbn      = {978-0813349107}
}

@article{beg2022,
  author = {Beg, Marijan and Lang, Martin and Fangohr, Hans},
  journal = {IEEE Transactions on Magnetics},
  title = {Ubermag: Towards more effective micromagnetic workflows},
  year = {2022},
  volume = {58},
  number = {2},
  pages = {1-5},
  doi = {10.1109/TMAG.2021.3078896}
}

@Misc{Cortes2019,
  author       = {David Cort{\'e}s-Ortu{\~n}o},
  title        = {OOMMFPy},
  howpublished = {Zenodo doi:10.5281/zenodo.2611194. Github: https://github.com/davidcortesortuno/oommfpy},
  year         = {2019},
  doi          = {10.5281/zenodo.2611194},
  url          = {https://doi.org/10.5281/zenodo.2611194},
}

@manual{cuda2024,
  title        = {{CUDA Toolkit Documentation}},
  author       = {{NVIDIA Corporation}},
  organization = {NVIDIA},
  year         = {2024},
  note         = {\url{https://docs.nvidia.com/cuda/}},
}

@book{chicone2006ordinary,
  title     = {Ordinary Differential Equations with Applications},
  author    = {Chicone, Carmen},
  year      = {2006},
  publisher = {Springer},
  address   = {New York, NY},
  series    = {Texts in Applied Mathematics},
  volume    = {34},
  edition   = {2nd},
  isbn      = {978-0387307695}
}
\end{document}